# Frequency Stability Transfer in Passive Mode-Locked Quantum-dash Laser Diode using Optical Injection Locking

K. Manamanni, T. Steshchenko, F. Wiotte, A. Chaouche Ramdane, M-O. Sahni, V. Roncin
and F. Du-Burck.

*Abstract*— In this paper, we present an experimental study of the metrological stabilization of a solid-state frequency comb for embedded metrology applications. The comb is a passively mode-locked laser diode based on InGaAs/InP Quantum-dash structure emitting optical lines into a 9 nm bandwidth centered at 1.55 µm with a repetition rate of 10.09 GHz. The frequency stabilization is achieved by optical injection locking of the comb with an external cavity laser diode referenced onto a metrological frequency standard. One observes the transfer of the spectral purity from the injection laser to the neighbouring modes of the injected one as well as the transfer of stability to the adjacent modes. The measurement of the long term stability highlights a frequency noise with random walk behavior specific of the passive mode locking process. Demonstration of sidebands of the injection laser at the repetition frequency of the comb also makes it possible to propose a transfer mechanism and to consider a complete stabilization of the frequency comb at a metrological stability level.

*Index Terms*—Semiconductor lasers, Mode-locked lasers, Frequency combs, Optical Injection Locking.

## I. Introduction

THIS paper presents experimental results on the frequency stability transfer from an external cavity laser diode (ECLD) stabilized onto a metrological reference towards a passive mode-locked laser diode (MLLD), thanks to optical injection locking [1]. The use of compact setups for stabilizing such semiconductor based frequency comb sources is of particular interest for metrological applications requiring a compact frequency reference either in the optical or in the microwave domain or both. A transportable measuring system for fundamental physics and spectroscopy or clocks comparison is already reported [2]. Small size semiconductor lasers are used for high resolution and sensitive measurements of gas spectra in the Mid-IR range [3]. High repetition frequencies in the 10 GHz range facilitate the stability transfer from optical to Radio-Frequency domain (RF) for cold atoms interrogation in atomic fountain frequency standards [4].

Such optical frequency combs could be used for the synthesis of low noise and long term stable microwave and THz frequencies for radar and deep space navigation [5]. The maturity of epitaxial technologies (molecular beam epitaxy or metal-organic chemical vapor phase epitaxy) for InAs/GaAs growth on InP or GaAs substrate allows 3D confined quantum dot structures. Ridge structures based on one or two sections, with or without saturable absorber section, in Fabry-Perot or distributed feedback configurations give passive MLLD generating self-pulsating light. Some of those lasers offer interesting characteristics such as high nonlinear properties induced by the material dimensionality and a low confinement factor able to give low amplitude and phase noise level [6]. Such passive MLLD structures emitting at 1.3 µm and 1.5 µm have been intensively investigated principally for their ability to generate both short optical pulses at high repetition rates and wide frequency combs [7]. At 1.55 µm, InGaAs quantum-dash structures offer single section lasers that generate sub-picosecond optical pulses width, jitter performance of a few hundred femtoseconds and repetition frequency rate higher than 100 GHz [8]. The frequency stabilization of such compact semiconductor frequency combs may be achieved by stabilizing both the optical frequency of one mode of the comb and the repetition rate of the optical pulses which is equivalent to the stabilization of the frequency shift between the comb teeth $f_{rep}$. Like fiber based frequency combs [9], Fabry-Perot semiconductor mode locked lasers present dispersive material that implies the generation of chirped optical pulse, but without internal compensation

Manuscript received November 1, 2021. This work was supported by the LABEX Cluster of Excellence FIRST-TF (ANR-10-LABX-48-01), within the Program "Investissements d'Avenir" operated by the French National Research Agency (ANR).
All Authors are with LPL « Laboratoire de Physique des Lasers », CNRS UMR-7538, Université Sorbonne Paris Nord, 99 av. J.-B. Clément, 93430 Villetaneuse, FR (corresponding author to provide phone: +33-149-403-246; fax: +33-149-403-200; e-mail: vincent.roncin@univ-paris13.fr).

possibility. Another difference is that taking into account the monolithic structure of their cavity the optical modes and $f_{rep}$ are coupled since the intra-cavity optical index and the cavity length are inextricably linked through the optical length.

An exhaustive review on the MLLD has mentioned three main techniques able to significantly reduce the $f_{rep}$ linewidth and the timing jitter of the optical pulses generated by the laser [10]: hybrid mode-locking by modulating the carrier density with an external RF source; optoelectronic feedback by directly modulating the carrier density with the detected beat-note at $f_{rep}$ and dual-mode optical injection using a single mode laser AM modulated at $f_{rep}$. Linewidth reduction in the entire comb has been also achieved using a feedforward technique based on the detection and correction of the frequency noise common to all the optical modes, using an electro-optic modulator located at the laser's output [11].

For the purpose of frequency comb stabilization, we describe here the implementation of a low-power optical injection locking technique. Optical injection offers advantage of a wide locking bandwidth at high injection power level, given the possibility to both transfer the spectral purity of the injection laser and compensate the frequency drift of the free-running laser. However, for a high injection power level, the counterpart of the wide locking bandwidth is a drastic reduction of the comb bandwidth [12]. Therefore, a low injection power is an interesting alternative for the frequency lock between the injected laser (MLLD) and the injection laser (ECLD) while preserving the entire bandwidth of the frequency comb [13]. We demonstrate that the comb frequency stabilization may be achieved at low injection power. This leads to a continuous operation over long time scale by combining passive insulation of the lasers and the use of ultra-low noise current sources [14].

In our setup, the stability of the ECLD frequency stabilized onto a metrological reference is transferred to an optical mode of the MLLD laser by low power optical injection. The long-term stability of all the modes of the comb can be characterized, given the possibility to study the mechanism of stability transfer from the injected mode to the other modes of the comb as detailed hereafter. The setup used for the frequency transfer is described in Section 2. It is based on a fiber ring-cavity used for transferring the stability from the metrological reference at 1542 nm to the injection laser at 1553 nm. Section 3 presents the frequency stability results achieved by optical injection locking. Finally, additional experimental characterization results are given in Section 4, making realistic assumptions about the stability transfer mechanisms in order to understand and overcome the limitations observed in the stability transfer.

II. EXPERIMENTAL DESCRIPTIONS

A. MLLD characteristics

The MLLD is based on a laser enclosed into a module fabricated by III-V lab [15]. The QD-based hetero-structure has been grown thanks to gas source molecular beam epitaxy (GSMBE) on a S-doped (100) InP substrate. The active core consists of 6 InAs QDashs layers stacked within GaInAsP injector quantum well emitting at 1.55 µm. The ridge waveguide laser was processed by a combination of dry and wet etchings. The facets were cleaved to get a 4.5 mm single section Fabry-Perot laser (FP). The FP chip was integrated into a butterfly module containing a temperature probe, a Peltier cooler and a microwave V-type microwave connector for high frequency current modulation. The device was initially developed for all-optical clock recovery in the framework of high bit-rate optical fiber communications [16]. For this purpose, the laser operating as an optical RF oscillator at a frequency $f_{rep}$ corresponding to its free spectral range is optically phase-locked by optical injection of on-off keying data stream presenting an intense clock line [17].

For a bias current around 200 mA at room temperature the FP laser modes are phase-locked within the 9 nm bandwidth which corresponds to a frequency comb of 125 lines spaced by $f_{rep}$ = 10.09 GHz and centered at 1551 nm with a flat spectrum as presented in Fig. 1. The use of a low-noise current source gives optical linewidths limited by the pulse-to-pulse jitter within a 6-35 MHz range [18]. The total optical power at the butterfly package fiber's output is 5 mW which corresponds to -19 dBm per mode. The mode-locking process results in a coherent beat-note of the comb lines at $f_{rep}$ with a linewidth in the kHz range. The sensitivity with the temperature of the laser diode is 10 GHz/K and an active control gives a stability better than 0.2mK until 100 s.

B. Experimental set-up for stability transfer

The scheme used for the frequency stabilization of the comb is presented in Fig. 1. Its three main parts are first the metrological reference at 1542 nm, then the fiber ring-cavity that transfers the frequency stability of the reference to a laser at 1550 nm and finally the MLLD which is optically injected by this stable laser.

The metrological reference used for the stability transfer is provided by LNE-SYRTE (Observatoire de Paris) through the French metrological network REFIMEVE+. This is a single-frequency Erbium-doped fiber-laser emitting at 1542.13 nm (channel 44 of WDM ITU-T) stabilized onto an ultra-stable Fabry-Perot cavity [19]. The fractional stability of this reference is better than $1.10^{-15}$ between 1 s and $10^5$ s of integration time.

The setup used for both the stabilization of the comb and the characterization of the stability transfer is detailed in Fig. 2. The fiber ring cavity stabilized onto the reference is used for transferring the stability towards the frequency comb. The stability transfer is achieved by frequency locking the tunable ECLD ($T_1$) to a cavity resonance close to the mode of the MLLD optically injected by $T_1$. The characterization of the stability transfer along the comb is achieved using a second tunable laser ($T_2$) which is also locked onto another resonance of the cavity. The beat-note obtained between $T_2$ and the comb mode is detected with an amplified high speed photodetector ensuring an analysis bandwidth of 12 GHz suitable for the wideband optical injection monitoring.

The detailed operating principle of the fiber cavity was described in a previous work [20]. The new fiber ring-cavity used in this work is composed of 200 m of single mode optical fiber (SMF) and a 10/90 optical coupler. We obtain a free spectral range (FSR) of 1 MHz and a resonance linewidth of 23 kHz resulting in a finesse of 43 and a quality factor of $8.4 \times 10^9$. The cavity length is controlled by 25 turns of optical fiber coiled around a 7.6 cm diameter PZT ring actuator with a diameter of 7.6 cm. The long-term stability is achieved by an active thermal control of the fiber with a Peltier module on which, 16% of the fiber length is laid. The whole cavity is enclosed into a homemade wood box improved for acoustic and thermal sensitivity reduction. Polarization controllers are located ahead of the cavity for ensuring the selection of one eigen polarization mode of the cavity and maintaining the long term polarization stability in the set-up.

For the frequency lock of the cavity, a phase modulation is applied to the reference at $f_3$ = 240 kHz and the Pound-Drever-Hall (PDH) signal is detected by a photodetector. A homemade electronics (demodulation and proportional-integral (PI) controller) drives the PZT ring actuator and the Peltier module respectively for fast and slow control of the cavity length.

The two tunable lasers are identical ECLDs. Each one is locked onto the fiber cavity. Both are driven by custom designed low-noise current sources ($\sigma_{Irms}$<1µA) allowing to get optical linewidths lower than 100 kHz. Both ECLD are modulated in two EOM's respectively at $f_1$ = 1.7 MHz and $f_2$ = 450 kHz and with modulation indexes of ~1. They are injected into the stable fiber cavity and detected after the optical circulator. The PDH signals at each frequency are detected by demodulation circuits followed by PI controllers. The control signals are added to the current of each ECLDs. The frequency stabilization leads to a significant reduction of their optical linewidths with a minimum linewidth $\Delta\nu_T$~200 Hz for $T_1$. The stability of the cavity is thus transferred to each laser ($T_1$ and $T_2$) but electronics discrepancy between the two locks leads to different linewidths ($\Delta\nu_T$~2 kHz for $T_2$) and stability levels. Thus, the fractional stability of the first ECLD as compared to the metrological reference at 1542 nm is in the $10^{-15}$ range for 1s integration time whereas the fractional stability level of the beat-note between $T_1$ and $T_2$ is in the $10^{-14}$ range. The stability transfer along the cavity has been measured without any significant degradation until 30 GHz. These results will be detailed soon in an article dedicated to the frequency transfer cavity.

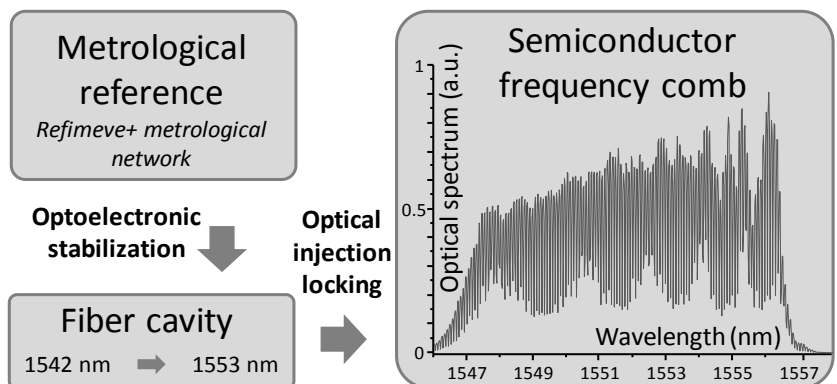

*Fig. 1. Scheme of the stability transfer from de metrological reference to the semiconductor frequency comb.*

Optical injection into the MLLD is achieved with $T_1$ locked to the cavity (Fig.2) in order to transfer its stability and its spectral purity. First, $T_1$ is injected into one mode of the MLLD. The wavelength is mechanically adjusted over 100 nm by turning a screw for rotating the internal diffraction gratings of the ECLD. A more precise adjustment is obtained with the current. Once the MLLD is optically injected, $T_1$ is locked to the cavity.

The characterization of MLLD comb lines is achieved through the beat-note between the comb spectrum and the second tunable laser $T_2$ which is shifted close to one particular mode. In order to reach metrological stability and to be able to perform stability measurements, we also lock $T_2$ onto the cavity. The beat-note frequency is adjusted to the frequency counter input range, at 60 MHz corresponding to the center of a 5 MHz band-pass filter ahead of the counter.

The described setup allows optical injection of a MLLD with a high spectral purity and long-term frequency stability laser. Moreover, the use of a second frequency stabilized ECLD provide the stability characterization of all the comb modes.

III. EXPERIMENTAL RESULTS ON THE STABILITY TRANSFER

A. Characterization in free running

The optical mode linewidth evolution of the free running MLLD throughout the comb follows a parabolic trend [21]:

$$\Delta \nu_n = \Delta \nu_{n_0}^{min} + (n - n_0)^2 \times \Delta f_{rep} \qquad (1)$$

Where $n_0$ is the index of the mode with the minimum linewidth $\Delta \nu_{n_0}^{min}$ and $\Delta f_{rep}$ the linewidth of the beat note of all optical modes detected in RF domain at $f_{rep}$ [22]. The spectrum of $f_{rep}$ shown in Fig. 3 (solid line) gives a full width at half maximum (FWHM) linewidth of 2 kHz. Expression (1) is established by assuming a white frequency noise for $f_{rep}$ with a power spectral density $S_{f_{rep}}$, which leads to a random walk process of the relative phase fluctuations between two adjacent lines in the comb and to a Lorentzian line shape of the RF beat note spectrum with a FWHM linewidth of $\pi \times S_{f_{rep}}$. This white noise is then dominant in a passive mode-locked laser for time scale smaller than the correlation time of the mode-locking process. But for a time scale longer than the correlation time, more advanced analyses consider noise processes in $1/f^\alpha$ with $\alpha \sim 1$, which becomes dominant and can be observed on the pulse train as a slow pulse shape breathing [23]. Such a process can also be studied through the long-time stability of the comb lines, as we will do in the following sections. Our experiment allows long time observation of $\nu_n$ and $f_{rep}$ and then, to observe non-stationary frequency noise such as frequency drift and random walk. These kinds of noise will be characterized through the Allan deviation. If the narrowest mode has a white noise of power density $S_{\delta \nu_{n_0}}$, its line shape is Lorentzian with a FWHM of $\pi \times S_{\delta \nu_{n_0}}$.

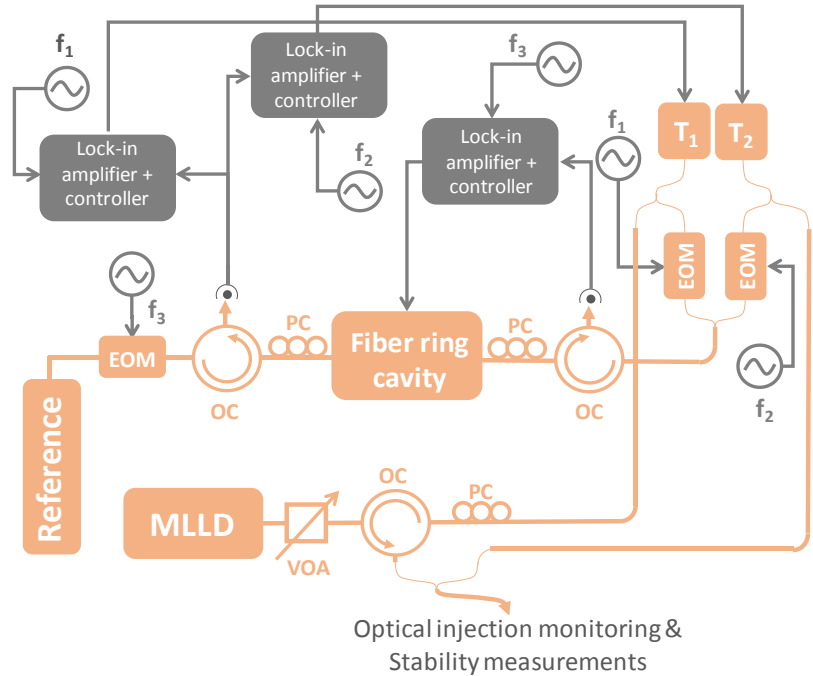

Fig. 2. Set-up of the MLLD stabilization by optical injection locking of $T_1$ in one mode of the comb: OC for Optical Circulator, PC for Polarization Controller, VOA for Variable Optical Attenuator and EOM for Electro Optic Modulator. Orange lines represent optical fibers and grey lines represent electrical cables.

The addition of technical noise results in a broadening and distortion of the mode (and from (1) of all modes). However, there is almost no change in the RF beat spectrum, reflecting the strong correlation between the modes induced by the passive mode-locking process. Consequently, the simple description of the comb given by (1) makes it possible to study the degrees of freedom of the comb $\nu_{n_0}$ and $f_{rep}$, as well as their sensitivity to parameters affecting the cavity optical length such as current and temperature fluctuations.

In Fig.4, we measure the mode linewidth $\Delta \nu_n$ as a function of the mode wavelength for three different currents. These measurements are deduced from the beat-note between $T_2$, which is frequency locked on a mode of the transfer cavity as explained in the previous section, and the modes of the free running MLLD. In agreement with (1), a parabolic evolution is obtained. For the three different currents, one observes a variation of the minimum linewidth $\Delta \nu_{n_0}^{min}$, a shift of the corresponding frequencies $\nu_{n_0}$, and a variation of the parabola concavity. According to (1), the values of $\nu_{n_0}$, $\Delta \nu_{n_0}^{min}$ and $\Delta f_{rep}$ are deduced for each current from a parabolic fit of the experimental data. It is found that the FWHM linewidth $\Delta f_{rep}$ decreases with the current with a slope of -3.7 Hz/mA, leading to a mitigation of the parabola concavity, [11], [12], [24]. This value is consistent with the direct linewidth measurements performed using the FWHM linewidth of a Lorentzian fit, that gives a slope of -3,1 Hz/mA. On the other hand, $\Delta v_{n_0}^{min}$ also decreases with the current following a slope of -26 kHz/mA but for the highest current value of 375 mA, the minimum linewidth is degraded by the technical noise of the current source [25]. Only taking into account the quantum noise in the laser (amplified spontaneous emission noise), the linewidth $\Delta f_{rep}$ and $\Delta v_{n_0}$ are correlated within a ratio of 7000.

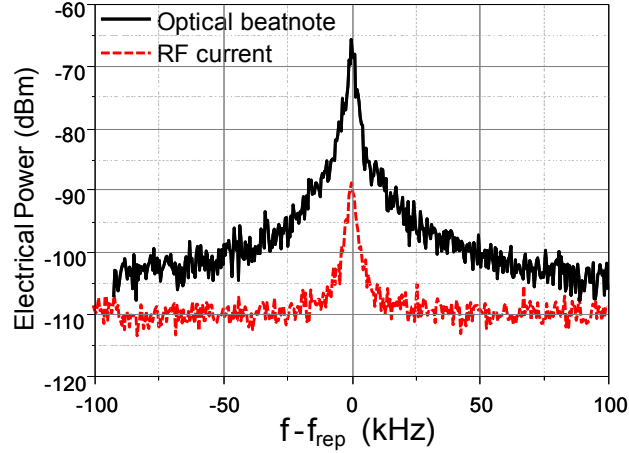

Fig. 3. Spectra at $f_{rep}$ : from the beat-note of all the MLLD modes into a high-speed photodetector of 12.5 GHz bandwidth (solid-trace), and from the current directly plugged at the RF SMA connector (dashed-trace).

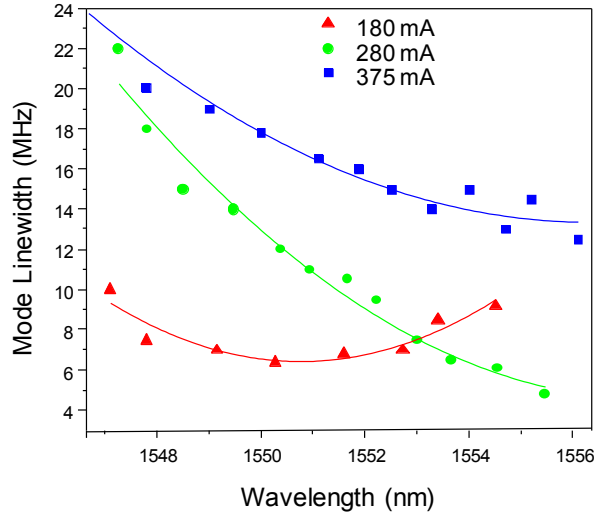

Fig. 4. Sensitivity of the comb mode linewidth $\Delta v$ with the current.

### B. Stability transfer via optical injection locking

In our experiment, a mode of the comb is optically injected with the laser $T_1$ frequency-locked to a mode of the transfer cavity. Its spectral purity is $\Delta v_{T_1} = 200$ Hz and its stability is in the $10^{-14}$ range. The frequency fluctuations of the injected mode copy those of the laser $T_1$. As a result, both the spectral purity of $T_1$ and its long-term stability are transferred to the injected mode. The frequency of the injected mode is:

$$v_N = f_{CEO} + N \times f_{rep} \qquad (2)$$

with $f_{CEO}$ the "Carrier Envelop Offset", $N$ the index of the injected comb line and $f_{rep}$ the frequency difference between two successive modes. Due to the optical injection, the injected mode becomes the narrowest one of the comb with a linewidth $\Delta v_N = \Delta v_{T_1}$. The stability of the different modes of the MLLD is measured by shifting $T_2$ along the comb and performing the beat with the studied mode. Taking the injected mode as reference, the other modes of the comb are identified by $n$ such as:

$$v_{N+n} = v_N + n \times f_{rep} \qquad (3)$$

In the following N+n / $T_2$ represents the beat-note between the comb mode number N+n and the stable laser $T_2$ used for the stability analysis.

The investigation of the frequency comb stabilization via optical injection with a frequency-stabilized laser is performed with a current of 200 mA giving a spectrum of 9 nm width centered at 1551 nm. The narrowest mode linewidth (5 MHz) is found at 1551 nm. Injection wavelength of $T_1$ is about 1553 nm corresponding to a mode linewidth of 7 MHz. Injection power of $T_1$ is -19 dBm at the VOA's output (cf. Fig. 2), in order to avoid spectrum distortion or narrowing caused by strong optical injection level [21]. Note that the frequency locking of laser $T_1$ onto the cavity leads to a stable optical injection despite a narrow injection bandwidth of 20 MHz.

The experimental results of the MLLD frequency stabilization *via* the optical injection with $T_1$ are reported in Fig. 5. Fig. 5 (a) shows the spectrum of several modes adjacent to the injected one at 1553 nm obtained from their beat-note with $T_2$. For $n = 1$ (at $\nu_N \pm f_{rep}$), the spectrum shows sidebands at 100 kHz characteristic of the locking bandwidth of the ECLD locked onto the fiber cavity, already observed on the beat-note $T_1 / T_2$. It is also observed for $n = 2, 3$ and 5 that the linewidth increases as one moves away from the injected mode from 2 to 60 kHz. For $n = 10$, the linewidth is estimated to ~500 kHz. Fig.5 (b) shows the stabilities in relative values (*i.e.* normalized to the 193 THz optical frequency) of the optical modes deduced from the Allan deviation of their beat note with $T_2$. The red circles correspond to the beat-note $T_1 / T_2$ when the lasers are shifted by 60 MHz from each other and both locked onto the fiber cavity. The Allan deviation, which is the square root of the quadratic sum of the stabilities of each laser, gives the sensitivity of our stability measurements. A characteristic trend of $3.10^{-13} \times \tau^{-1/2}$ is observed until 20 seconds which is attributed to a dominant white frequency noise leading to a linewidth of 2.4 kHz [26]. This value is consistent with the linewidth measurement of the beat-note $T_1 / T_2$. For integration times above 50 s, an evolution in $\tau^1$ is observed, characteristic of a linear frequency drift. With this result, we consider that the use of the beat-note $T_1 / T_2$ limits the performance of our stability characterization to that of $T_2$ locked onto the transfer cavity.

The stability of an optical mode at 1551 nm of the free running MLLD is in right-triangles. A trend in $\tau^0$ is observed for integration time until 10 s and one in $\tau^{+1/2}$ for higher values of the integration time. As we shall see hereafter, this $\tau^{+1/2}$ trend appears systematically in long term stability measurements of the optical modes of this laser, whether injected or not. To our knowledge, this characteristic behavior of a random walk frequency has never been reported for semiconductor MLLD. In temperature controlled laser diodes, the influence of an external temperature drift on the butterfly mount may lead to a frequency drift in $\tau^{+1}$. No temperature variation showing a trend in $\tau^{+1/2}$ was observed in our set-up.

Let us emphasize that we were able to highlight this behavior owing to the long-term stability of both the injection and analysis

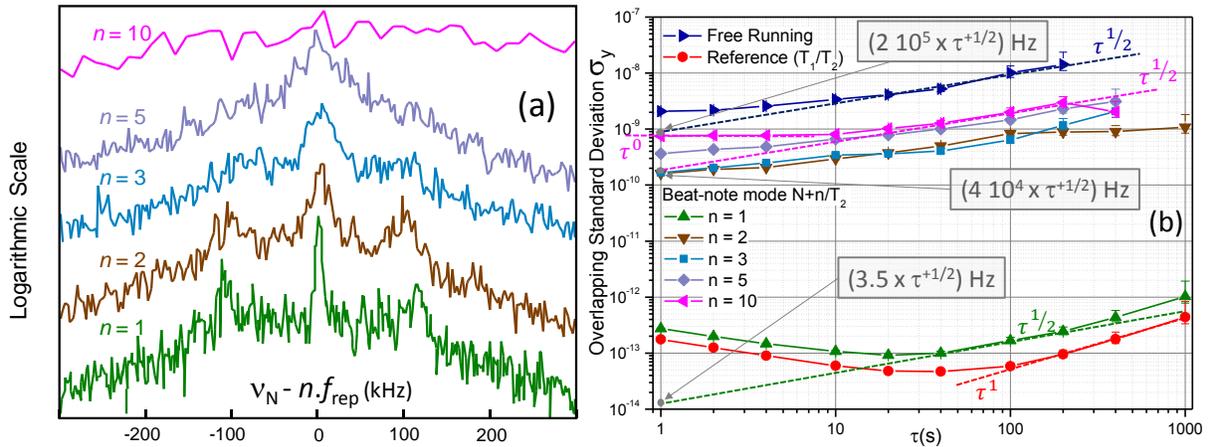

Fig. 5. Stability transfer from the injected laser $T_1$ to the optical modes of the MLLD: (a) the spectral purity transfer to the adjacent modes N+n considering the harmonic numbers n =1, 2, 3,5 and 10, and (b) the long term

laser. In particular, the frequency stability of the mode in free running reaches 2 MHz after 100 s, well below the injection locking bandwidth of 20 MHz, which allows for stable injection up to 1000 s.

The stability of the optical modes $\nu_{N+n}$ is then measured when the mode $N$ is optically injected by $T_1$. Our setup does not allow direct measurement of the injected mode stability since the beat-note is dazzled by the reflection of $T_1$ at the MLLD input facet. Thus, we measure the stability of the beat-notes N+n / $T_2$ with n > 0. The curve with up-triangles in Fig. 5 (b) is the beat-note N+1 / $T_2$ and corresponds to the square root of the quadratic sum of the frequency fluctuations of the laser $T_1$ and the mode N+1. The stability is very close to that of the reference $T_1 / T_2$ and demonstrates an efficient stability transfer from $T_1$ towards the adjacent modes of the injected one. A trend in $\tau^{+1}$, characteristic of a linear frequency drift, dominates for integration times greater than 200 s. Nevertheless, it can be noted that an additional trend in $(3.5 \times \tau^{+1/2})$ Hz in absolute value is observed from 30 to 200 s. It is noticeable that the stabilities N+2 / $T_2$ (down-triangles), N+3 / $T_2$ (squares) and N+5 / $T_2$

(diamonds) are considerably degraded with respect to that of $N+1/T_2$, with a value at 1 s between 2 and $5.10^{-10}$ against $3.10^{-13}$ for $N+1 / T_2$. On the other hand, it is noted that in each case, a $\tau^{+1/2}$ stability evolution corresponding to a random walk frequency is clearly observed in the long term. Finally, the stability of the beat-note $N+10 / T_2$ (left-triangles) shows a threshold at $(2\ 10^5 \times \tau^0)$ Hz and an evolution given by $(4\ 10^4 \times \tau^{+1/2})$ Hz, i.e. a $\tau^{+1/2}$ trend common to all measurements, except to that of the stability reference. The second degree of freedom of the comb is the repetition rate $f_{rep}$. Its stability measurement is achieved through the beat-note at $f_{rep}$ measured on a fast photodiode and down-shifted into the band of the frequency counter using a high frequency mixer and a 10 GHz local oscillator. Fig. 6 shows the stability of $f_{rep}$ for the free running MLLD (squares), for the MLLD under optical injection in the mode $N$ (triangles) and for the free-running MLLD with a 1 dBm electrical modulation at $f_{rep}$ delivered by an external signal generator directly plugged to the SMA connector of the MLLD (circles). The first observation is that, without electrical modulation, the stability of $f_{rep}$ show no significant difference with (triangles) and without (squares) optical injection. According to the linewidth of $f_{rep}$ for a Lorentzian shape estimated at about 2 kHz, it is expected for short integration times that the frequency stability is dominated by a white frequency noise in $(25 \times \tau^{-1/2})$ Hz. Actually, one can observe that the stability until 10 seconds is dominated by a technical noise generating a slow frequency modulation (presence of a bump at 2 s) and for integration times higher than 10 seconds one observes a random walk frequency in $(120 \times \tau^{+1/2})$ Hz. This trend has already been observed in Fig. 5 for the optical modes, as well as for the MLLD in both in free running and optically injected. From these observations, we may consider that the frequency stabilities evolving in $\tau^{+1/2}$ are common for both $\nu_n$ and $f_{rep}$. The ratio between $\sigma_{frep}$ and $\sigma_{\nu_n}$ is $2.10^5 /120 = 1666$.

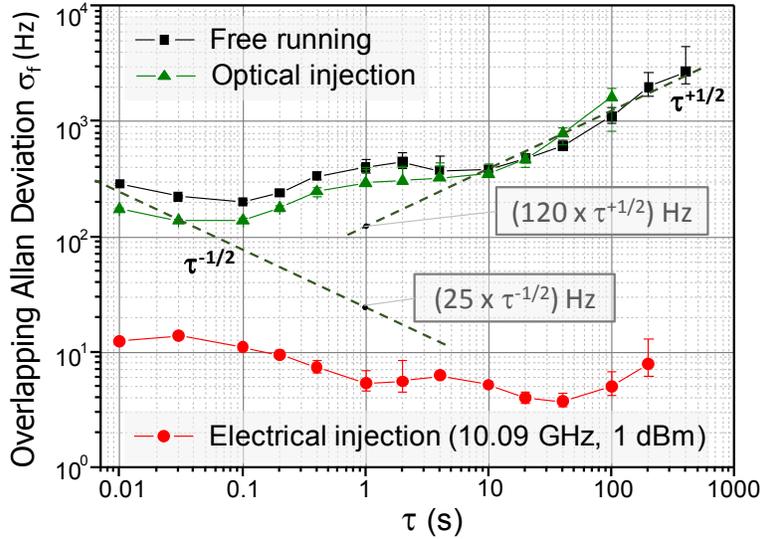

Fig. 6. Frequency stability results of the RF beat-note at $f_{rep}$ resulting from the beating of all MLLD modes into a high-speed photodetector of 12 GHz bandwidth.

The curve with circles in Fig. 6 corresponds to the stability of $f_{rep}$ for the free-running MLLD with an electrical modulation. This external modulation of the current at $f_{rep}$ increases its stability and cancels the slope in $\tau^{+1/2}$. Despite this, it should be noted that we did not succeed in efficient optical injection locking and current modulation at $f_{rep}$. simultaneously This is attributed to the phase difference between the external modulation and the internal mode-locking conditions based on a constant phase relationship between all the optical modes [24]. To overcome this problem, one can think to modulate directly the current with the beat-note between two optical modes of the optically injected comb.

IV. ANALYSIS AND DISCUSSION

In this work, we have shown the noise transfer in MLLD lasers. First, the frequency white noise of the optical modes that is transferred to $f_{rep}$ was observed from the optical mode widths and the $f_{rep}$ width. Then, thanks to our device for characterizing the stability of the modes and $f_{rep}$, we have also observed the transfer of a random walk frequency noise in the comb, never mentioned in the literature to our knowledge. In this section, we propose a physical origin of this random walk frequency observed in MLLD lasers followed by an analysis of the stability transfer process by optical injection. This last point aims at understanding why the stability transfer is limited to the mode immediately adjacent to the injected one.

## A. Origin of the frequency random walk

The frequency random walk is observed over all Allan deviations, both for the optical modes and for the RF beat-note whether the laser is optically injected or not. Only when the current is modulated by an external source does this frequency random walk disappear, suggesting that this behavior is characteristic of the passive mode-locking process.

Although there is no complete theoretical description of the mode locking process in single section passively mode locked lasers, it is accepted that mode-locking arises from the gain non-linearity of the medium. In particular, the four-wave mixing (FWM) phenomenon plays a key role [27]. The FWM causes mutual injection of the modes leading to a correlation of the phases between the laser modes, at the cause of the passive mode locking. The consequence is that the linewidth of the intensity spectrum lines (the RF spectrum) resulting from the beat between the optical modes are considerably narrower (~kHz) than the optical modes themselves (a few MHz).

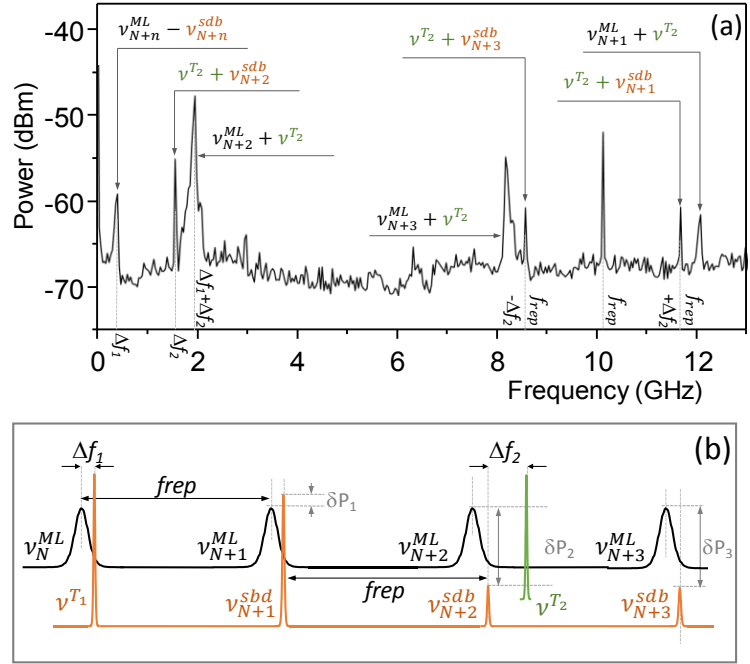

Fig. 7. Beat-note between the analysis laser $T_2$ and the MLLD output spectrum injected with $T_1$ detuned out of the locking bandwidth (a). Sketch summarizing the analysis of the peaks in the spectrum and demonstrating the presence of sidebands from $T_1$ generated within the injection process (b).

It should be noted that in ordinary multimode Fabry-Perot lasers the material dispersion imposes an evolution of the free spectral range (FSR) along the spectrum [28], [29]. In the case of discussed mode-locked FP laser, FWM process creates sidebands at $f_{rep}$ frequency that optically inject the neighboring modes and manage the locking of the whole comb with a strictly constant frequency shift $f_{rep}$ between the modes. In that way, FWM adjusts an initially inhomogeneous FSR induced by material dispersion to a common value $f_{rep}$. It is known that the injection locking process of a slave oscillator by a master oscillator imposes a relationship between their phase relations under injection, and their frequency relations without injection [30]. This is easily transposable to optical oscillators and demonstrate that under optical injection, the phase of the master laser manages the frequency of the injected laser If we consider that the white frequency noise of optical modes is due to spontaneous emission, it is admitted that the optical phase undergoes a random walk. Moreover, it has been shown that in the case of passive MLLD, the differential phase between adjacent modes (i.e. the phase fluctuations of $f_{rep}$) also undergoes a random walk that leads to a non-stationary process [24].

Following the hypothesis of a mode locking process which is managed by FWM in such MLLD, it is therefore expected that this random walk of the phase will lead to a correlated random walk of the frequency of the injected modes and consequently a random walk of $f_{rep}$.

## B. Process of stability transfer along the comb

For analyzing the stability results presented above and understanding the mechanisms involved in the stability transfer from the laser $T_1$ to the semiconductor frequency comb, we propose further experimental investigations of the injection locking process. For this purpose, the resulting signal from the MLLD injected by $T_1$ beating with $T_2$ is shown in Fig. 7 (a). Here, $T_1$ is outside the injection bandwidth and the MLLD is not locked. A sketch is proposed in Fig. 7 (b) to help the analysis of this spectrum. The writing conventions are the following: $\nu_N^{ML}$ is the injected mode, $\nu_{N+n}^{ML}$ the mode frequency shifted by $n \times f_{rep}$ from the injected one; $\nu_N^{T_1}$ the injection laser $T_1$ injected into the mode $N$ and $\nu^{T_2}$, the analysis laser $T_2$. In the configuration presented in Fig. 7 (a), $T_2$ is shifted by $2 \times f_{rep} + \Delta f_2$ from $T_1$, the latter being slightly detuned by $\Delta f_1$ from $\nu_N^{ML}$ according to the sketch of Fig. 7 (b).

We note at once that the spectrum in Fig. 7 (a) exhibits both narrow and broad lines. The narrow lines at $f_{rep}$ results from all the beat-notes of adjacent modes pairs in the MLLD spectrum. Three additional narrow lines appear in the spectrum of Fig. 7 (a) which are necessarily due to the injection of $T_1$ into the MLLD, thus resulting in a set of sidebands generated in the laser active medium and shifted by $f_{rep}$ from $T_1$. These sidebands are denoted by $\nu_{N+n}^{sbd}$ with $n = 1, 2$ or $3$. They lead to the three narrow lines in the spectrum of Fig. 7 (a) resulting from their beats with the laser $T_2$ at the frequencies $\Delta f_2$ for $\nu_{N+2}^{sbd}$, $f_{rep} - \Delta f_2$ for $\nu_{N+3}^{sbd}$ and $f_{rep} + \Delta f_2$ for $\nu_{N+1}^{sbd}$. The broad lines result from the beat-notes between the optical modes of the comb (~5MHz) and the analysis laser $T_2$ or the sidebands of $T_1$.

The spectrum of Fig. 7 (a) also gives an estimate of the powers in each sidebands $\nu_{N+n}^{sbd}$, considering the powers of the modes of MLLD and $T_2$ to be constant. We consider 125 modes within the MLLD spectrum of 5 mW, a coupling coefficient between the laser and the output fiber of 0.3, and deduce that each mode has a power of about 12 µW. Thus, the power of each sideband is: $P\nu_{N+1}^{sbd} = 17$ µW; $P\nu_{N+2}^{sbd} = 2.2$ µW and $P\nu_{N+3}^{sbd} = 2.6$ µW. As mentioned above, the broad lines result from the beat-notes between a comb mode $\nu_{N+n}^{ML}$ and $T_1$ sidebands or $T_2$. The first peak at $\Delta f_1$ comes from the set of beat-notes between $\nu_{N+n}^{ML}$ and $T_1$ sidebands $\nu_{N+n}^{sbd}$ ($n = 1, 2, 3, …$). The broader peak at $\Delta f_1 + \Delta f_2$ is the beat-note between $\nu_{N+2}^{ML}$ and $\nu^{T_2}$. Finally, the peak close to the frequency $f_{rep} - \Delta f_2$ results from the beat-note between $T_2$ and the mode $n = 3$ and that corresponding to the beat-note between $T_2$ and the mode $n = 1$ is close to the frequency $f_{rep} + \Delta f_2$.

In order to understand the origin of the sidebands, we can refer to Fig. 3 giving the spectrum of current at $f_{rep}$ of the free-running MLLD (dashed trace). A spectral component with a linewidth of ~1 kHz is observed demonstrating the modulation of the carrier density, and thus of the gain at $f_{rep}$. According to the phase-amplitude coupling in such MLLD (linewidth enhancement factor), we expect that any laser injected into the MLLD undergoes both amplitude modulation (AM) and phase modulation (PM) at frequency $f_{rep}$. Moreover, FWM efficiency in such component could also accounts for the presence of high order sidebands and consequently participates in the multiple injection process in the MLLD.

The observation of such sidebands which may have various origins (AM, FM, FWM) during the optical injection of the MLLD suggests that they could be involved in the transfer of spectral purity and stability from the injected mode to the neighbouring modes. The degradation of spectral purity and stability observed on the modes further away from the injected mode ($n > 1$) can be related to the decrease of sideband power with $n$. These first hypotheses will of course have to be supported by further experimental and theoretical studies. In particular, it will be necessary to understand why we observe a progressive degradation for the spectral purity with *n* and a sudden change for the stability.

V. CONCLUSION

We present in this article a study of a Fabry-Perot QDash-based laser diode for metrological applications and particularly for compact solutions. The MLLD generates a frequency comb thanks to an efficient passive mode locking process. We have described the first experimental results obtained with an original setup allowing the frequency comb stabilization by optical injection with an ultra-stable ECLD.
The current and temperature stabilization of the MLLD ensures stable injection at low optical power that avoids distorting the MLLD emission spectrum. The impact of the optical injection on the comb modes is analysed from the beat-note between the studied modes and a second ultra-stable ECLD.
The first key result of our study is the systematic existence of a random walk frequency noise of the comb modes, with or without optical injection of the MLLD. We have shown that the relative frequency noise between adjacent modes is also dominated by a random walk process. This effect seems to be characteristic of the passive mode-locking process since it disappears in the case of an active mode-locking (electrical modulation of the $f_{rep}$).
The second important result is that while the optical injection allows the transfer of spectral purity from the injection laser to the injected mode and neighbouring modes with a progressive degradation for the more distant modes, the transfer of long-term stability is effective only for the injected mode and the immediately adjacent ones. The stability of the $N + n$ modes with $n > 1$ is only slightly improved as compared to the one of the free laser.
We have also shown that the interaction of the injected laser with the gain medium of the MLLD leads to $f_{rep}$-shifted sidebands, which can be associated with the generation of a component at $f_{rep}$ in the MLLD supply current, demonstrating the existence of a modulation processes at this frequency.

We have formulated some first assumptions to explain the origin of our observations. In particular, the frequency random walk could be linked to the passive locking process within the comb by mutual injection of adjacent optical modes through a FWM mechanism. On the other hand, our observations suggest that the frequency stability transfer from the injected mode to the adjacent modes of the comb is related to the sidebands resulting from the interaction of the injected laser with the MLLD.

MLLD-based frequency combs are an attractive solution for the realization of compact metrological devices. It is essential to understand the mechanisms underlying the transfer of spectral purity and stability within the comb. The characterization device presented here allowed an original study of the long-term stability of the optical modes of the MLLD with or without injection. To our knowledge, the presented results have never been reported in the literature.

ACKNOWLEDGMENT

We warmly thank Yenista Optics for providing the ECLD sources, Morgan Advanced Materials for piezoelectric tubes used in the transfer cavity, A. Shen and F. Van-Dijk from GIE III-V Lab for providing the QD-MLLD sources and CNRS FOTON UMR6082 for fruitful scientific interactions. We also thank O. Lopez, E. Cantin and A. Amy-Klein at LPL and P.-E. Pottie and R. Letargat at LNE-SYRTE for providing the REFIMEVE+ metrological signal. We finally thank H. Mouhamad and L. Malinge for their daily electronics support.